\documentclass[12pt]{article}
\usepackage{sigsam, amsmath,xspace,listings,color}

\usepackage[colorinlistoftodos]{todonotes}

\issue{appeared in Volume 57, Issue 2}
\articlehead{RISC Report Series 23-11 
}
\titlehead{Computing Mellin representations and asymptotics: the RICA package}
\authorhead{J. Bl\"umlein, N. Fadeev, C. Schneider}
\newcommand{\dint}[4]{\ensuremath{\int_{#1}^{#2}\mathrm{d}#3\, #4}\xspace}

\definecolor{blaugrau}{rgb}{0.796887, 0.789075, 0.871107}

\newcounter{mmacnt}
\def\restartmma{\setcounter{mmacnt}{0}}
\restartmma \catcode`|=\active
\def|#1|{\mathrm{#1}}
\catcode`|=12
\newenvironment{mma}{
	\par
	\catcode`|=\active
	\parskip=2pt\parindent=0pt 
	\small
	\def\In##1\\{%
		\def\linebreak{\hfill\break\null\qquad}%
		\refstepcounter{mmacnt}
		\hangindent=2.5em\hangafter=0
		\leavevmode
		\llap{\tiny\sffamily In[\arabic{mmacnt}]:=\kern.5em}%
		\mathversion{bold}\footnotesize$\tt\bf\displaystyle##1$\normalsize
		\mathversion{normal}\par
	}%
	\def\Print##1\\{%
		\def\linebreak{\hfill\break}%
		\hangindent=2.5em\hangafter=0
		\leavevmode\scriptsize ##1\par}%
	\def\Out##1\\{%
		\vspace*{-0.2cm}\def\linebreak{$\hfill\break\null\hfill$}%
		\kern\abovedisplayskip\par
		\hangindent=2.5em\hangafter=0
		\leavevmode
		\llap{\tiny\sffamily Out[\arabic{mmacnt}]=\kern.5em}
		\footnotesize$\displaystyle\tt##1$\normalsize\hfill\null\par
		\kern\belowdisplayskip\vspace*{-0.3cm}
	}%
	\def\Warning##1##2\\{%
		\def\linebreak{\hfill\break}%
		\hangindent=2.5em\hangafter=0
		\leavevmode
		{\scriptsize##1 : ##2}\par}%
}{%
	\par\smallskip
}

\newcommand{\LoadP}[1]{\fcolorbox{black}{blaugrau}{
				\begin{minipage}[t]{15cm}
					\footnotesize #1
		\end{minipage}}}

\begin{document}
\title{Computing Mellin representations and asymptotics of nested binomial sums in a 
symbolic way: the RICA package\footnote{Supported by the Austrian Science Foundation (FWF) grant P33530.}}
\author{Johannes Bl\"umlein$^a$, Nikolai Fadeev$^b$, Carsten Schneider$^b$ \\
$^a$ Deutsches Elektronen-Synchrotron, DESY, 15738 Zeuthen, Germany \\ 
$^b$ Research Institute for Symbolic Computation (RISC)\\
Johannes Kepler University Linz, A-4040 Linz, Austria\\
{\footnotesize \url{Johannes.Bluemlein@desy.de, Nikolai.Fadeev@risc.jku.at, Carsten.Schneider@risc.jku.at}}}

\maketitle

\begin{abstract}
Nested binomial sums form a particular class of sums that arise in the context of particle physics computations at higher orders in perturbation theory within QCD and QED, but that are also mathematically relevant, e.g., in combinatorics. We present the package RICA (Rule Induced Convolutions for Asymptotics), which aims at calculating Mellin representations and asymptotic expansions at infinity of those objects. These representations are of particular interest to perform analytic continuations of such sums. 
\end{abstract}

The package RICA, which stands for \textbf{R}ule \textbf{I}nduced \textbf{C}onvolutions 
for \textbf{A}symptotics, stem from the need to deal in a systematic way with finite and 
infinite nested binomial and inverse binomial 
sums that appear in the context of particle physics computations at higher orders in 
perturbation theory within QCD and QED \cite{ABSal:2014}.  
This kind of analytic calculations of Feynman integrals, aim at high precision predictions 
for particle physics experiments for processes with massless partons but also massive quarks in 
QCD. They involve on the mathematical side the computation of increasingly complex 
iterated integrals.
In the process of doing so, many different classes of functions arise, for example (generalized) 
harmonic sums and (generalized) harmonic polylogarithms \cite{V:1999}--\cite{ABS:2013}.  The 
package \verb!HarmonicSums!, developed by Jakob Ablinger \cite{A:2012}, allows to manipulate 
efficiently such objects, and in particular find closed forms or simpler representations, compute 
(inverse) Mellin transforms, asymptotics, etc.\ thereof. Nevertheless, extensions of nested 
generalized harmonic sums weighted by the binomial coefficient $\binom{2n}{n}$ cannot be 
dealt with in the general case, and a different approach to treat them has been presented in 
\cite{ABRS:2014}.  Instead of relying on Mellin inversion through solving of differential equations 
\cite{A:2016},  the method described in the former paper relies on recursive computations of 
Mellin convolutions of the summands of the iterated binomial sums. The main advantage of this 
approach is that by defining a set of possible general cases,  i.e., classes of functions for 
which we can compute in general the Mellin inverse and/or the Mellin convolution,  the computation 
of the Mellin inverse can be made rather fast and straightforward. Moreover, by identifying new 
general cases and adding them to our ``dictionary'', we can easily extend the classes of iterated 
sums we can deal with, even beyond the binomial case.

Another advantage in representing the nested binomial sums as Mellin integrals is that one can 
perform asymptotic expansions at infinity,  which helps us to obtain for example analytic continuations 
of those nested sums \cite{ABRS:2014}.  There exist several possible methods to perform asymptotic 
expansions of functions of the form $\dint{0}{1}{x}{x^n f(x)}$, i.e., defined as Mellin transforms of 
some function $f$, depending on the regularity of $f$.  In particular, there is a rather general method 
\cite{N:1906,L:1906} that relies on changes of variables and term by term integration 
\cite{ABRS:2014}, which has been implemented in the \verb!RICA! package together with several 
variations or simpler methods to speed up computations. 

\medskip

Before presenting the main functions and an example of what our package can do, we want to 
emphasize the fact that while \verb!RICA! implements and extends 
algorithms and methods that 
have not been implemented yet, it strongly relies on notations, objects and tools provided 
both by Carsten Schneider's \verb!Sigma! package \cite{S:2007,S:ToApp} and Jakob Ablinger's 
\verb!HarmonicSums! \cite{A:2012} package.

Let us define some notations,  following \cite{ABRS:2014}. We consider binomial nested sums, 
i.e. nested sum of the form:
$$BS(n)=\sum_{i_1=1}^n a_1(i_1;b_1,c_1,m_1)\sum_{i_2=1}^n a_2(i_2;b_2,c_2,m_2)\cdots 
\sum_{i_k=1}^n a_k(i_k;b_k,c_k,m_k)$$
where $a_p(i;b,c,m)=\binom{2i}{i}^b \frac{c^i}{i^m},\ b\in\{-1,0,1\},\ c\in \mathbb{R}^\star,\ m\in\mathbb{N}$, and almost all of 
the $b$ are $0$ except for some that fit into the case of 
the theorems described in~\cite[Section 4]{ABRS:2014}.  We also accept coefficients that have 
a more general structure, such as $\frac{c^i}{2i+1}\binom{2i}{i}^{b}$, $b\in\{-1,1\}$, that we extended to cases of 
the form $\frac{p(i)}{q(i)}\binom{2i}{i}$, where $p,q\in \mathbb{C}[X]$, $\deg p \leq \deg 
q$.\footnote{Currently we are working on extensions that cover any rational function and  
more complicated coefficients.} 

\medskip

Now we present the main functionalities of our package. Once installed, we have to load it in 
Mathematica:

\begin{mma}
\In <<RICA.m;\\
\Print \LoadP{Rule Induced Convolutions for Asymptotics (RICA) package by Nikolai Fadeev \copyright\ RISC-JKU}\\
\end{mma}

\medskip

\noindent As stated before,  \verb!RICA! depends on \verb!Sigma! and \verb!HarmonicSums! which have to be loaded beforehand. 

The three main functions of the package are the following:
\begin{itemize}
\item \verb!SumToMellin[expr,c,x,opts]!: Given an argument  \verb!expr! that is a linear 
combination of binomial nested sums, it computes its Mellin inverse.  More precisely, the argument 
\verb!expr! has the following form:

$$\mathrm{expr}=\sum_{i=1}^m\alpha_i \mathrm{GS}\left[ \{f_{i,1}(\mathrm{VarGL}), 
f_{i,2}(\mathrm{VarGL}),\ldots, f_{i,p_i}(\mathrm{VarGL})\},n \right],\alpha_i\in\mathbb{C},\ 
p_i\in\mathbb{N}$$

and \verb!GS! is the generalized iterated sum defined in \verb!HarmonicSums! as
$$\mathrm{GS}\left[ \{f_{1}(\mathrm{VarGL}), f_{2}(\mathrm{VarGL}),\ldots, f_{p}(\mathrm{VarGL})\},n 
\right]=\sum_{1\leq i_p\leq i_{p-1}\leq \cdots \leq i_1\leq n}f_1(i_1)f_2(i_2)\cdots f_p(i_p).$$

The second argument \verb!c! specifies a variable that is used to denote 
the constants that might appear in the final expression\footnote{In particular if we give for example 
$C$, the constants will be denoted as $C_i$.},  and $x$ is the integration variable in the Mellin 
integrals.  Finally, opts is an optional boolean argument called \verb!ToGLbBasis! which uses 
\verb!HarmonicSums! basis reduction functions to simplify the result further; its default value is 
\verb!True!.

The output is returned in the form of a tuple where the first element is a linear combination of 
Mellin transforms, using \verb!HarmonicSums! notation $\mathrm{Mellin}[k_n(x),f(x)]=\int_0^1 
\mathrm{d}x\, k_n(x)f(x)$ where $k_n(x)=(ax)^n$ or $k_n(x)=(ax)^n-1$ is the kernel,  i.e.
$$\mathrm{expr}=\sum_{i=1}^p \alpha_i\,\mathrm{Mellin}[k_{i,n}(x),f_i(x)].$$

The second element is then a list of  the values of those constants, i.e., 
$$\{C_1\rightarrow \mathrm{value}_1,\ldots, C_p \rightarrow \mathrm{value}_p\}$$
that might arise inside of the functions $f_i(x)$.

\item \verb!AsymptoticsMellint[expr, x, n, ord, opts]!: This function computes the asymptotic 
expansion of a linear combination of Mellin integrals.  It takes as the main argument \verb!expr! 
which is a linear combination of Mellin representations as obtained in the output of \verb!SumToMellin!, together with 
\verb!x! which is the Mellin integration variable, \verb!n! which is the Mellin parameter,  \verb!ord! which is the required order 
of the expansion, and \verb!opts! which allows to insert possible fine-tuning options.

The output is returned as an asymptotic expansion similar to what Mathematica does when using the 
built-in function \verb!Series!.

\item \verb!AsymptoticsSum[expr,n,x,ord]!: The function can be seen as a ``combination'' of both functions 
above, i.e. ``\verb!AsymptoticsSums = AsymptoticsMellint !$\circ$\verb! SumToMellin!''.  It takes 
therefore as the main argument \verb!expr!, i.e. a linear combination of nested binomial sums in the 
\verb!GS! representation, together with the Mellin parameter \verb!n!, Mellin integration variable \verb!x! and desired expansion order \verb!ord!. 

The output is in the same form as \verb!AsymptoticsMellint!.
\end{itemize}

To conclude, below is an explicit example of a computation using all functions above.  We consider 
the nested binomial sum:
$$S(n)=\sum_{i=1}^n \frac{1}{i\binom{2i}{i}}\sum_{j=1}^i (-2)^j \tbinom{2j}{j} = GS\left[\left\{\frac{1}{
\mathrm{VarGL}\tbinom{2\mathrm{VarGL}}{\mathrm{VarGL}}},(-2)^\mathrm{VarGL} \tbinom{2\mathrm{VarGL}}{
\mathrm{VarGL}}\right\},n\right]$$
by introducing it in Mathematica as follows:

\begin{mma}
\In  sum1=GS\left[\left\{\frac{1}{\text{VarGL}*\text{Binomial}(2 \text{VarGL},\text{VarGL})},(-2)^{\text{VarGL}}* \text{Binomial}(2 \text{VarGL},\text{VarGL})\right\},n\right]; \\
\end{mma}

\noindent Now we apply \verb!SumToMellin! to the input expression and get:

\begin{mma}
\In mel1=SumToMellin[sum1,C,x]\\
\Out \left\{Mellin\left[(-2)^n x^n-1,\frac{1-\frac{1}{6 \sqrt{2} \sqrt{x+\frac{1}{8}}}}{x+\frac{1}{2}}\right]-\frac{2}{3} \text{Mellin}\left[4^{-n} x^n-1,\frac{1}{\sqrt{1-x} (x-4)}\right],\{\}\right\} \\
\end{mma}

\noindent As we see,  the result is a sum of two Mellin integral representations where no extra constants arise.  We are now in the position to apply the function \verb!AsymptoticsMellint! to get
the asymptotic expansion, e.g., up to order 4:

\begin{mma}
\In AsymptoticsMellint[mel1[[1]],x,n,7]\\
\Out -\frac{26425 (-1)^n 2^{n+3}}{531441 n^7}-\frac{2213 (-1)^n 2^{n+3}}{177147 n^6}+\frac{799 (-1)^n 2^{n+3}}{59049 n^5}+\frac{73 (-1)^n 2^{n+3}}{6561 n^4}-\frac{5 (-1)^n 2^{n+3}}{729 n^3}-\frac{7 (-1)^n 2^{n+3}}{243 n^2}+\frac{(-1)^n 2^{n+4}}{27 n}\\
\end{mma}

Using the function \verb!AsymptoticsSum!, if we only wanted the asymptotic expansion and not the 
Mellin representation we could have simply called it directly:

\begin{mma}
\In AsymptoticsSum[sum1,n,x,7]\\
\Out -\frac{26425 (-1)^n 2^{n+3}}{531441 n^7}-\frac{2213 (-1)^n 2^{n+3}}{177147 n^6}+\frac{799 (-1)^n 2^{n+3}}{59049 n^5}+\frac{73 (-1)^n 2^{n+3}}{6561 n^4}-\frac{5 (-1)^n 2^{n+3}}{729 n^3}-\frac{7 (-1)^n 2^{n+3}}{243 n^2}+\frac{(-1)^n 2^{n+4}}{27 n}\\
\end{mma}

\end{document}